\documentstyle[12pt,a4]{article}

\begin{document}

\baselineskip=17pt plus 0.2pt minus 0.1pt


\makeatletter

\def\CN{{\cal N}}
\def\d{{\rm d}}
\def\Tr{\mathop{\rm Tr}}
\def\L{{\rm L}}
\def\T{{\rm T}}
\def\calL{{\cal L}}
\def\calD{{\cal D}}
\def\calG{{\cal G}}
\def\calH{{\cal H}}
\def\del{{\partial}}
\def\half{{1 \over 2}}
\def\nn{ \nonumber }
\newcommand{\dlt}{\delta_\epsilon}
\newcommand{\Dlt}{\Delta_\epsilon}
\newcommand{\bra}[1]{\left\langle #1\right|}
\newcommand{\ket}[1]{\left| #1\right\rangle}
\newcommand{\VEV}[1]{\left\langle #1\right\rangle}
\newcommand{\braket}[2]{\VEV{#1 | #2}}
\newcommand{\diag}{\mathop{\rm diag}}
\newcommand{\bb}{\!\!\!}
\newcommand{\Deta}{\Delta^{(\eta)}}
\newcommand{\dltot}{\widetilde{\delta}_\epsilon}
\newcommand{\Goneloop}{\Gamma_{1\hbox{\scriptsize -loop}}}
\newdimen\ex@
\ex@.2326ex
\def\dddot#1{\raise\ex@\hbox{${\mathop{#1}\limits^{
             \vbox to-1.4\ex@{\kern-\tw@\ex@\hbox{\rm...}\vss}}}$}}
\def\ddddot#1{\raise\ex@\hbox{${\mathop{#1}\limits^{
             \vbox to-1.4\ex@{\kern-\tw@\ex@\hbox{\rm....}\vss}}}$}}

\makeatother


\begin{titlepage}
\title{
\hfill\parbox{4cm}
{\normalsize KUNS-1553\\{\tt hep-th/9901034}}\\
\vspace{1cm}
Conformal Symmetry in Matrix Model\\
beyond the Eikonal Approximation
}
\author{
Hiroyuki {\sc Hata}\thanks{{\tt hata@gauge.scphys.kyoto-u.ac.jp}\ \
Supported in part by Grant-in-Aid for Scientific
Research from Ministry of Education, Science and Culture
(\#09640346).}
{} and
Sanefumi {\sc Moriyama}
\thanks{{\tt moriyama@gauge.scphys.kyoto-u.ac.jp}}
\\[7pt]
{\it Department of Physics, Kyoto University, Kyoto 606-8502, Japan}
}
\date{\normalsize January, 1999}
\maketitle
\thispagestyle{empty}

\begin{abstract}
\normalsize

Recently, it was shown by Jevicki, Kazama and Yoneya (JKY)
that the Super-Yang-Mills theories (SYM) in $D\leq 4$ can
reproduce the conformal symmetry of the near-horizon geometry of the
D-brane solutions.
However, the eikonal approximation they used is not sufficient for
analyzing the conformal symmetry in SYM.
We carry out the 1-loop calculation beyond the eikonal approximation
for $D=1$ SYM (i.e., Matrix model) to confirm the claim of JKY.

\end{abstract}

\end{titlepage}

\section{Introduction}
The duality between string theory on the Anti-de-Sitter (AdS) space
and the conformal field theory on its boundary has been investigated
intensively \cite{Malda,GKPW}.
Recently, an interesting observation was made by Jevicki, Kazama and
Yoneya (JKY) \cite{JKY3} that the isometry of the AdS space can be
derived from $D=4$ $\CN=4$ Super-Yang-Mills theory (SYM).
The special conformal transformation (SCT) in SYM is modified by a
quantum extra term to reproduce the isometry of the AdS space.

Although the near-horizon geometry of the D$p$-brane solution with
$p\ne 3$ is not of the AdS type, SCT as well as the
dilatation symmetry\footnote{
The SCT and the dilatation symmetry in both the near-horizon geometry
with $p\ne 3$ and the SYM with $D\ne 4$ are generalized ones in the
sense that they transform also the coupling constants \cite{JKY0}.}
of the near-horizon geometry play important roles in determining the
form of the Born-Infeld action of the D$p$-brane \cite{Malda,JY}.
The SCT of the near-horizon geometry differs from that in SYM
by a $U$ (radial coordinate) dependent term which vanishes on the
boundary $U\to\infty$.
JKY \cite{JKY0} generalized the arguments of \cite{JKY3} to the
$p\ne 3$ case to show that this $U$-dependent term can be reproduced
from the extra term in the SCT of SYM which we have to add for
compensating the breaking of SCT due to the gauge-fixing term.

In \cite{JKY0}, they calculated the SCT and the 1-loop effective
action using the eikonal approximation and examined the SCT Ward
identity in SYM, in particular for the $D=1$ case.
However, the eikonal approximation is in fact insufficient for
determining the effective action and examining the SCT Ward identity.
This is because the higher derivative terms of the scalar fields
(transverse coordinates of the D-brane), which vanish in the eikonal
approximation, are converted into terms containing only the first
derivative of the scalars through the integration by parts.

The purpose of this note is to carry out the 1-loop calculation beyond
the eikonal approximation for $D=1$ SYM
(i.e., Matrix model \cite{BFSS}).  We determine
completely the form of the 1-loop effective action with a particular
operator dimension and confirm the SCT Ward identity.

\section{SCT in Matrix model}

First, let us briefly summarize the SCT in Matrix model
\cite{JY,JKY3,JKY0}.
It is well known that the action of $D=4$ $\CN=4$ SYM has
conformal symmetry.
This is also the case for maximally SYM in other dimensions
if we assign the coupling constant a conformal dimension and
vary it under the transformation.
We will here concentrate on the $D=1$ SYM case (in the Euclidean
formulation):
\begin{equation}
S_{\rm SYM} = \int\! \d \tau\,\Tr \left(
{1 \over 2 g_s} (D X_m)^2
-{1 \over 4 g_s} [X_m,X_n]^2
-\half \theta^T D \theta
-\half \theta^T \gamma^m [ X_m,\theta ] \right) .
\label{SYM}
\end{equation}
This is the action of D-particles and is familiar in the Matrix model.
The action (\ref{SYM}) is invariant under both the dilatation
and SCT. In particular, the transformation law of the bosonic
variables under SCT reads
\begin{eqnarray}
\dlt \tau\bb &=&\bb\epsilon \tau^2 , \nn\\
\dlt (A, X_m)\bb &=&\bb -2\epsilon \tau (A, X_m) ,\label{SCT}\\
\dlt g_s\bb &=&\bb -6\epsilon \tau g_s .\nn
\end{eqnarray}
In order to quantize the system, we have to add the gauge-fixing and
the corresponding ghost terms to the original action (\ref{SYM}):
\begin{eqnarray}
&&S_{\rm gf} = \int {\d \tau \over 2 g_s} \Tr G^2 , \qquad
G = -\del A + i [B_m,Y_m]  , \nn\\
&&S_{\rm gh} = i\int \d \tau \Tr\left( \bar{C}\del D C
 - \bar{C}[B_m,[X_m,C]] \right) .
\label{gf+gh}
\end{eqnarray}
Here we have adopted the background gauge, and $Y_m$ is the
fluctuation of the scalars $X_m$ from the diagonal background $B_m$;
$X_m = B_m + Y_m$.
While $S_{\rm gf}+S_{\rm gh}$ (\ref{gf+gh}) is dilatation invariant,
it is not invariant under SCT (\ref{SCT}).
However, the SCT symmetry is restored by applying a BRST
transformation, $\delta_{\rm BRST} X_m = i[X_m,C] \lambda$, etc.\
with a field dependent transformation parameter $\lambda$,
\begin{eqnarray}
\lambda = -2i \int \d\tau\Tr\left(
\bar{C}(\tau) A(\tau)\,\epsilon\right).
\end{eqnarray}
In fact, the violation of SCT,
$\dlt\!\left(S_{\rm gf}+S_{\rm gh}\right)$,
is canceled by the change of the functional measures of $A$ and
$\bar{C}$ under $\delta_{\rm BRST}$.
Therefore, the effective action $\Gamma[B,g_s]$ of the present system
satisfies the following SCT Ward identity,
\begin{equation}
\int\!\d\tau\left(\dlt g_s(\tau)\frac{\delta}{\delta g_s(\tau)}
+\left(\dlt + \Dlt\right)\! B_{m,i}(\tau)\,
\frac{\delta}{\delta B_{m,i}(\tau)}
\right)\Gamma[B,g_s]=0 ,
\label{WI}
\end{equation}
where the extra term $\Dlt B_{m,i}$ for
$B_m=\diag\left(B_{m,i}\right)$ is
\begin{eqnarray}
\Dlt B_{m,i}(\tau)= -2 \VEV{ [C(\tau),X_m(\tau)]_{ii}
\int\! \d\tau_0\,\Tr (\bar{C}(\tau_0) A(\tau_0)\,\epsilon ) }  .
\label{extraterm}
\end{eqnarray}
Note that the total SCT for $B_m$ is now given as a sum of the
classical part $\dlt B_{m,i}=-2\epsilon\tau B_{m,i}$ and the quantum
correction $\Dlt B_{m,i}$.

In the following sections, we shall first calculate explicitly
$\Gamma[B,g_s]$ and $\Dlt B_{m,i}$ within the eikonal approximation
and find that the eikonal approximation is insufficient for
confirming the SCT Ward identity (\ref{WI}). We then present
a systematic analysis beyond the eikonal approximation by
incorporating the acceleration and the higher time derivative terms.

\section{Eikonal approximation}

In this section we shall repeat the calculation of $\Gamma[B,g_s]$ and
$\Dlt B_{m,i}$ in the eikonal approximation given in \cite{JKY0}
in a more systematic manner to clarify the problem.
Before carrying out the calculation, let
us recall the structure of the double expansion in the Matrix model.
According to \cite{BBPT}, calculations in the Matrix model
has the structure of the double expansion with respect to
$[\L^{-3}]$ and $[\L^{-2}\T^{-2}]$,
where $[\L]$ represents the length dimension,
and $[\T]$ the time dimension of the operators.
The expansion with respect to $[\L^{-3}]$ corresponds to
the loop expansion, while $[\L^{-2}\T^{-2}]$ is the unit dimension of
the time derivative expansion. Therefore, the $n$-loop effective
action $\Gamma_{n\hbox{\scriptsize -loop}}$ has the following structure:
\begin{equation}
\begin{array}{c c c c c c c c c}
\Gamma_{0\hbox{\scriptsize -loop}}
           &=& [\L^2\T^{-2}] &+& [\L^0\T^{-4}]
           &+& [\L^{-2}\T^{-6}] &+& \cdots \\
\Goneloop  &=& [\L^{-1}\T^{-2}] &+& [\L^{-3}\T^{-4}]
           &+& [\L^{-5}\T^{-6}] &+& \cdots \\
\Gamma_{2\hbox{\scriptsize -loop}}
           &=& [\L^{-4}\T^{-2}] &+& [\L^{-6}\T^{-4}]
           &+& [\L^{-8}\T^{-6}] &+& \cdots \\
&& && \vdots && &&
\end{array}
\label{expansion}
\end{equation}
Here we are interested in the $[\L^2\T^{-2}]$ term of
$\Gamma_{0\hbox{\scriptsize -loop}}$, and the $[\L^{-1}\T^{-2}]$ and
$[\L^{-3}\T^{-4}]$ terms of $\Goneloop$.

For the 1-loop calculation we need the quadratic part of the
Matrix model action.
As stressed in \cite{JKY0}, it is necessary to take into account the
fact that the coupling constant $g_s$ is now a function of $\tau$ and
to keep the dependence linear in $\eta\equiv \dot g_s/g_s$:
\begin{eqnarray}
\calL_{YY}\bb &=&\bb {1\over 2g_s} Y_{m,ij}(-\del^2
 + \eta\del + B_{ij}^2)Y_{m,ji} , \\
\calL_{AA}\bb &=&\bb {1\over 2g_s} A_{ij}(-\del^2
 + \eta\del + B_{ij}^2)A_{ji} , \\
\calL_{YA}\bb &=&\bb {2i\over g_s} V_{m,ij} Y_{m,ij} A_{ji} , \\
\calL_{\theta \theta}\bb &=&\bb \half \theta_{\alpha ,ij}
( -\delta_{\alpha \beta} \del +\gamma^m_{\alpha \beta}
 B_{m,ij}) \theta_{\beta ,ji} , \\
\calL_{\bar{C} C}\bb &=&\bb - i\bar{C}_{ij}(-\del^2
 + B_{ij}^2)C_{ji} ,
\end{eqnarray}
where
\begin{eqnarray}
&&V_{m,ij}\equiv\dot{B}_{m,ij} - \half \eta B_{m,ij} , \\
&&B_{m,ij}\equiv B_{m,i}-B_{m,j} .
\end{eqnarray}

First, the 1-loop effective action is given by
\begin{eqnarray}
&&\Goneloop =
\Tr\Bigl\{ 10 \ln \left( -\del^2+\eta \del+B^2 \right)
      + \ln \bigl(1- 4 V_m \Deta V_m \Deta \bigr) \nn\\
&&\qquad\qquad
- 4 \ln \left( -\del^2+B^2+\dot{B} \right)
- 4 \ln \left( -\del^2+B^2-\dot{B} \right)
- 2 \ln \left( -\del ^2+B^2 \right) \Bigr\} ,
\label{one-loop}
\end{eqnarray}
with $\dot B\equiv\sqrt{(\dot B_m)^2}$.
Here, we are considering the two-body case with
$(B_{m,1},B_{m,2})=(0,B_m)$, and the propagator $\Deta$ is given by
$\Deta\equiv\left(-\del^2  + \eta\del + B^2\right)^{-1}$.
In (\ref{one-loop}), the first term is the contribution of
$\calL_{YY}$ and $\calL_{AA}$, the second term is due to the mixing
between $Y_m$ and $A$, the third and the fourth terms are from
$\calL_{\theta\theta}$, and the last term is the ghost loop.

Expanding (\ref{one-loop}) to ${\cal O}(\eta)$, we find that the
1-loop effective action is given as a sum of the following three
terms,
$\Goneloop=\Gamma_1+\Delta\Gamma_1+G_1$:
\begin{eqnarray}
\Gamma_1\bb &=&\bb\Tr \left\{ \ln \left( 1-4
\dot{B} \Delta \dot{B} \Delta \right)
-4 \ln \left( 1-
\dot{B} \Delta \dot{B} \Delta \right) \right\} , \label{Gamma1}\\
\Delta \Gamma_1\bb &=&\bb\Tr \left\{ 8\eta \del \Delta
+\eta \left( \del- \frac{\dot{B} \cdot B}{\dot{B}} \right)
\frac{1}{-\del^2+B^2+2\dot{B}} \right.\nn\\
&&\qquad\qquad \left.
+\eta \left( \del+ \frac{\dot{B} \cdot B}{\dot{B}} \right)
\frac{1}{-\del^2+B^2-2\dot{B}} \right\} , \label{DeltaGamma1}\\
G_1\bb &=&\bb\Tr
\left\{\ln \bigl( 1-4 V_m \Delta V_m \Delta \bigr)
-\ln \bigl( 1-4 V \Delta V \Delta \bigr) \right\} ,
\label{G1}
\end{eqnarray}
with $\Delta\equiv\left(-\del^2 + B^2\right)^{-1}$ and
$V\equiv\sqrt{(V_m)^2}$.
Here, $\Gamma_1$ is $\eta$-independent and $\Delta \Gamma_1$ is
proportional to $\eta$.
As for $G_1$, it vanishes in the eikonal approximation
although it makes important contributions in the next section.

In the calculation of the effective action in the eikonal
approximation,\footnote{Here we set $b\cdot v=0$.}
\begin{eqnarray}
B_m = b_m + v_m \tau , \label{eikonalB}
\end{eqnarray}
there appear the propagators
$\Delta_n\equiv\left(-\del^2 + b^2 + v^2 \tau^2 + n v\right)^{-1}$.
Its explicit expression is given by
\begin{eqnarray}
\bra{\tau_1}\Delta_n \ket{\tau_2}
\bb &=&\bb \int^\infty_0 \d s\, e^{-(b^2+n v) s}
\,\calD (\tau_1,\tau_2; s)  ,
\label{Delta_n}\\[5pt]
\calD (\tau_1,\tau_2; s)\bb &=&\bb\sqrt{v \over {2 \pi \sinh 2sv}}
\exp \left\{ -{v \over{2 \sinh 2sv}}
\left[\left(\tau_1^2+\tau_2^2\right)\cosh 2sv -2 \tau_1 \tau_2
\right] \right\} .
\end{eqnarray}
The following useful convolution formula holds for
$\calD(\tau_1,\tau_2;s)$:
\begin{eqnarray}
&&\calG (\tau_1,\tau_2;s_1,s_2;J)\equiv
\int\! d \tau_0\, \calD (\tau_1,\tau_0; s_1)\, e^{J\tau_0}\,
\calD (\tau_0,\tau_2;s_2) \nn\\
&&=\calD(\tau_1,\tau_2;s_1+s_2) \exp\left\{
\frac{J\left(
\tau_1\sinh 2 s_2 v + \tau_2\sinh 2 s_1 v\right)
+ J^2\sinh 2 s_1 v \sinh 2 s_2 v/2v}{\sinh 2(s_1+s_2)v}
\right\} .\nn\\
\label{convolution}
\end{eqnarray}
In particular, repeated use of (\ref{convolution}) gives
\begin{eqnarray}
\bra{\tau_1}\left(\Delta_0\right)^M\ket{\tau_2} = \int^\infty_0
\!\d s_1 \d s_2 \cdots \d s_M\,
e^{-b^2\sum_{k=1}^M s_k}
\calD (\tau_1,\tau_2; \sum_{k=1}^M s_k) .
\label{Delta_0^m}
\end{eqnarray}

Using the above formulas, the 1-loop effective action in the
eikonal approximation can be calculated exactly to any desired order
in the time derivative expansion (see the remarks below eq.\
(\ref{finalsct})).
Keeping the terms with dimensions $[\L^{-1}\T^{-2}]$ and
$[\L^{-3}\T^{-4}]$ as we mentioned at the beginning of this section,
we have
\begin{eqnarray}
\Gamma_1^{\rm eikonal}\bb &=&\bb -6 v^4\Tr\left(\Delta_0\right)^4
= \int\!\d \tau \left( -\frac{15}{16}
 \frac{v^4}{\rho^7} \right) ,\nn\\
\Delta \Gamma_1^{\rm eikonal}\bb &=&\bb
\Tr\biggl\{ 8 \eta \del \Delta_0
 +\sum_{\pm}\eta\left(\del\mp v\tau \right)\Delta_{\pm 2 }\biggr\}
=\int\!\d \tau\,\eta
\left( \frac{3}{2} \frac{v^2 \tau}{\rho^3} +\frac{15}{4}
 \frac{v^4 \tau}{\rho^7} -\frac{105}{16}
 \frac{v^6 \tau^3}{\rho^9} \right) ,\label{effective_1_eikonal_rho}\\
G_1^{\rm eikonal}\bb &=&\bb 0 , \nn
\end{eqnarray}
where $\rho^2\equiv b^2 + v^2\tau^2$ ($=B^2$ in the eikonal
approximation).

Note that the effective action should be a functional of $B_m(\tau)$
and $g_s(\tau)$ as is seen from the original expression
(\ref{one-loop}).
Then, our next task is to rewrite (\ref{effective_1_eikonal_rho})
in terms of $B_m(\tau)$ and its derivatives. In the eikonal
approximation with $\ddot B=0$, there are the following candidate
terms with suitable dimensions for expressing the above effective
actions:
\begin{eqnarray}
&&\frac{(\dot{B}^2)^2}{B^7}=\frac{v^4}{\rho^7} ,\quad
\frac{\dot{B}^2 (\dot{B} \cdot B)^2}{B^9}=\frac{v^6 \tau^2}{\rho^9} ,
\quad
\frac{(\dot{B} \cdot B)^4}{B^{11}}=\frac{v^8 \tau^4}{\rho^{11}} ,
\label{terms_Gamma1}\\
&&\frac{\dot{B} \cdot B}{B^3}=\frac{v^2 \tau}{\rho^3} ,\quad
\frac{\dot{B}^2 (\dot{B} \cdot B)}{B^7}=\frac{v^4 \tau}{\rho^7} ,\quad
\frac{(\dot{B} \cdot B)^3}{B^9}=\frac{v^6 \tau^3}{\rho^9} ,
\label{terms_DeltaGamma1}
\end{eqnarray}
where (\ref{terms_Gamma1}) is for $\Gamma_1$ and
(\ref{terms_DeltaGamma1}) for $\Delta\Gamma_1$.
Comparing (\ref{effective_1_eikonal_rho}) with (\ref{terms_Gamma1}) and
(\ref{terms_DeltaGamma1}), we find that
\begin{eqnarray}
\Gamma_1^{\rm eikonal}\bb &=&\bb \int\!\d \tau \left( -\frac{15}{16}
 \frac{(\dot{B}^2)^2}{B^7} \right) , \label{final_Gamma1_eikonal}\\
\Delta \Gamma_1^{\rm eikonal}\bb &=&\bb
 \int\!\d \tau\,\eta\left(
 \frac{3}{2}\frac{\dot{B} \cdot B}{B^3}
 +\frac{15}{4}\frac{\dot{B}^2 (\dot{B} \cdot B)}{B^7}
 -\frac{105}{16}\frac{(\dot{B} \cdot B)^3}{B^9} \right) .
\label{final_DeltaGamma1_eikonal}
\end{eqnarray}

Having finished the calculation of the effective action, let us
proceed to $\Dlt B_m$. Reviving temporarily the indices $(i,j)$ of the
fields, eq.\ (\ref{extraterm}) at the 1-loop order is given by
\begin{eqnarray}
\Dlt B_{m,i}(\tau) = -2 \int\!\d\tau_0  \left(
\langle C_{ij}(\tau)\,\bar{C}_{ji}(\tau_0)\rangle
\VEV{Y_{m,ji}(\tau)A_{ij}(\tau_0)\,\epsilon}
-(i\leftrightarrow j) \right) ,
\label{extraterm_oneloop}
\end{eqnarray}
where the free propagators are
\begin{eqnarray}
\langle C_{ij} (\tau_1) \bar{C}_{ij}(\tau_2)\rangle\bb &=&\bb
 i \bra{\tau_1}\Delta_{ij}\ket{\tau_2}  , \\
\langle Y_{m,ji}(\tau_1) A_{ij}(\tau_2)\rangle
\bb &=&\bb -2i g_s\bra{\tau_1}\Delta_{ij}\dot B_{m,ij}\Delta_{ij}
\left(1-4\dot B_{\ell,ij}\Delta_{ij}\dot B_{\ell,ij}\Delta_{ij}
\right)^{-1}\ket{\tau_2} ,
\end{eqnarray}
with $\Delta_{ij}\equiv\left(-\del^2 +B_{ij}^2\right)^{-1}$.

Returning again to the two-body situation and making the eikonal
approximation, eq.\ (\ref{extraterm_oneloop}) is reduced to
\begin{equation}
\Dlt B_{m}(\tau) = -8g_s\epsilon\bra{\tau}
\Delta_0 v_m\,\half\sum_{\pm}\Delta_{\pm 2}\Delta_0\ket{\tau}
=  -g_s\frac{3}{2} \frac{\epsilon v_m}{\rho^5} ,
\label{Dlt_B_final}
\end{equation}
where the final expression is obtained by using the formulas
(\ref{Delta_n})---(\ref{Delta_0^m}) and keeping only the lowest order
term $[\L^{-4}\T^{-1}]$ in the $[\L^{-2}\T^{-2}]$ expansion.
Expressing again (\ref{Dlt_B_final}) in terms of $B_m$, we
obtain the final form of the quantum SCT $\dltot\equiv\dlt+\Dlt$:
\begin{eqnarray}
&&\dltot \tau = \epsilon \tau^2 ,\qquad
\dltot g_s = -6\epsilon \tau g_s ,\nn\\
&&\dltot B_m =  -2\epsilon \tau B_m
-g_s\frac{3}{2} \frac{\epsilon \dot{B}_m}{B^5} .
\label{finalsct}
\end{eqnarray}
Although the coefficient $3/2$ of the last term of $\dltot B_m$ is
apparently different from that of the $U$-dependent term in the SCT of
the near-horizon geometry, the two SCT are related by a coordinate
transformation \cite{JKY0}.

Let us give remarks on the systematic expansion in powers of
$[\L^{-2}\T^{-2}]$. We shall explain the following two points:
\begin{enumerate}
\item
How to obtain the expressions
like eqs.\ (\ref{effective_1_eikonal_rho}) and (\ref{Dlt_B_final}),
which have the powers of $\rho$ (but not $b$) as their denominators.
\item
Note that the quantity $v^2\tau^2/\rho^2$ is dimensionless.
Hence, one might think that, for the terms with a given dimension,
an infinite series in powers of
$v^2\tau^2/\rho^2$ appears in the effective action and $\Dlt B_m$.
On the other hand, if these quantities can be expressed in terms of
$B_m$ and its derivatives, we should have only a finite number
of terms with a given dimension.
We shall explain why in the concrete calculations the series in powers
of $v^2\tau^2/\rho^2$ terminate at a finite order.
\end{enumerate}

To explain the first point, note that $\Goneloop$ and
$\Dlt B_m$ such as the second terms of (\ref{effective_1_eikonal_rho})
and (\ref{Dlt_B_final}) are given, after using the convolution
formulas (\ref{convolution}) and (\ref{Delta_0^m}), by the following
generic form:
\begin{equation}
\prod_{k=1}^M \int_0^\infty\! \d s_k\, F(s_1 v,\cdots,s_M v)
\calD(\tau,\tau;\sum_k s_k)
\,e^{-b^2\sum_k s_k} ,
\label{general_form}
\end{equation}
in terms of some function $F(s_k v)$. The expansion in powers of
$1/\rho$ (but not $1/b$) is obtained by rewriting the last two factors
in (\ref{general_form}) as
\begin{equation}
\calD(\tau,\tau;s)\,e^{-b^2 s}=
\sqrt{\frac{v}{2\pi\sinh 2sv}}\exp\left(-v\tau^2\left[
\tanh sv - sv\right]\right)\,e^{-\rho^2 s} ,
\label{keyformula}
\end{equation}
and Taylor-expanding the integrands in (\ref{general_form}) other than
$e^{-\rho^2\sum_k s_k}$ with respect to $v$.

The second point that the series in powers of $v^2\tau^2/\rho^2$
terminates at a finite order is easily understood from the series
expansion,
\begin{equation}
\exp\left(-v\tau^2\left[\tanh sv - sv\right]\right)
=\sum_{m=0}^{\infty} \sum_{n=0}^m c_{m,n}(sv)^{2m} (sv^2\tau^2)^n .
\end{equation}
An important point is that for a given $m$, which specifies the
dimension of the term in units of $[\L^{-1}\T^{-1}]$, the summation
over $n$ terminates at $n=m$.

Let us return to the results (\ref{final_Gamma1_eikonal}),
(\ref{final_DeltaGamma1_eikonal}) and (\ref{finalsct}) in the eikonal
approximation and examine whether the SCT Ward identity (\ref{WI})
holds. First, the effective action $\Gamma_0$ at the tree level is
given by
\begin{equation}
\Gamma_0 = \int \d \tau \frac{\dot{B}^2}{2 g_s} ,
\end{equation}
and it is transformed under SCT of (\ref{finalsct}) as
\begin{equation}
\dltot\Gamma_0 = \int \d \tau\,\frac{15}{4}
\frac{\epsilon\dot{B}^2(\dot{B} \cdot B)}{B^7} ,
\label{dltot_Gamma0}
\end{equation}
up to a total derivative term.
The claim of \cite{JKY0} is that the 1-loop effective action
$\Gamma_1 + \Delta\Gamma_1$ is given by
\begin{eqnarray}
\left(\Gamma_1 + \Delta\Gamma_1\right)^{\rm JKY}
= \int\!\d\tau\left\{-\frac{15}{16}\frac{(\dot{B}^2)^2}{B^7}
+ \eta(\tau )\left(\frac{3}{2}\frac{\dot B\cdot B}{B^3}
+ \frac{15}{8} \frac{\dot{B}^2 (\dot{B} \cdot B)}{B^7}
\right)\right\} .
\label{JKYclaim}
\end{eqnarray}
Then, since we have $\dltot\eta=-6\epsilon+{\cal O}(\eta)$, putting
$\eta=0$ after the transformation and discarding the total derivative
term $(3/2)(\dot B\cdot B)/B^3$, we have
\begin{eqnarray}
\dltot\left(\Gamma_1 + \Delta\Gamma_1\right)^{\rm JKY}
=\int \d \tau \left(- \frac{15}{4}
\frac{\epsilon\dot{B}^2(\dot{B} \cdot B)}{B^7} \right) ,
\end{eqnarray}
which just cancels (\ref{dltot_Gamma0}).
However, our $\Delta\Gamma_1$ (\ref{final_DeltaGamma1_eikonal}) does
not agree with the $\eta$ term in (\ref{JKYclaim}).

In the above argument, we have implicitly assumed that the expressions
(\ref{final_Gamma1_eikonal}), (\ref{final_DeltaGamma1_eikonal}) and
(\ref{finalsct}) are valid for a generic $B_m(\tau)$ not restricted
to the eikonal form (\ref{eikonalB}) although they are obtained using
the eikonal approximation.
However, if, for example, we added to
(\ref{final_DeltaGamma1_eikonal}) the term
$(15/8)(\dot B\cdot B)(\ddot B\cdot B)/B^7$, which vanishes
in the eikonal approximation, the sum of the last two terms in
(\ref{final_DeltaGamma1_eikonal}) and the added one would become
$(15/8)\dot{B}^2(\dot{B} \cdot B)/B^7
+(15/16)(\d/\d\tau)[(\dot{B}\cdot B)^2/B^7]$
with the desired coefficient $15/8$. On the other hand, if we further
added $-(3/4)(\ddot B\cdot\dot B)/B^5$ to the above quantity,
the sum would become a purely total derivative term.
Therefore, for examining the SCT Ward identity, we have to go beyond
the eikonal approximation. This is the subject of the next section.

\section{Beyond the eikonal approximation}

Before carrying out the calculation beyond the eikonal approximation,
let us list the terms we have to consider in the 1-loop effective
action $\Goneloop$ and the quantum SCT $\Dlt B_m$.
First, for $\Dlt B_m$, $\dot B_m/B^5$ is the unique function of $B_m$
with dimension $[\L^{-4}\T^{-1}]$. Therefore, the quantum SCT
(\ref{finalsct}) is already valid beyond the eikonal
approximation.
Next, let us consider the terms proportional to $\eta$ in $\Goneloop$
(they are contained in $G_1$ as well as in
$\Delta\Gamma_1$). We are interested in those terms with dimensions
$[\L^{-1}\T^{-1}]$ and $[\L^{-3}\T^{-3}]$ excluding $\eta$.
For $[\L^{-1}\T^{-1}]$, there is only one term $\dot{B}\cdot B/B^3$.
As for the terms with dimension $[\L^{-3}\T^{-3}]$,
we have $\dddot{B}\cdot B/B^5$, $\ddot{B}\cdot\dot{B}/B^5$ and
$(\ddot{B}\cdot B)(\dot{B}\cdot B)/B^7$, besides the terms
$\dot{B}^2(\dot{B}\cdot B)/B^7$ and $(\dot{B}\cdot B)^3/B^9$
which we have already taken into account in the eikonal approximation.
For the $\eta$-independent terms in $\Goneloop$, there are in
total 11 kinds of terms although we do not list them explicitly.
Our task in this section is to determine the coefficients of all of
these terms to examine the SCT Ward identity.

For this purpose, let us consider the following polynomial form for
the function $B_m(\tau)$:
\begin{eqnarray}
B_m(\tau) = b_m + v_m \tau + \half a_m \tau^2
+ \frac{1}{6} w_m \tau^3 .
\end{eqnarray}
and keep terms up to linear in $a\cdot b$, $a\cdot v$ and
$w\cdot b$ in $\Goneloop$. These three kinds of terms are sufficient
for determining all the coefficients in $\Goneloop$.
Then, the parts of the effective action proportional to $\eta$
and its derivatives are
\begin{eqnarray}
&&\Delta \Gamma_1 = \Delta \Gamma_1^{\rm eikonal}
+ \Tr\biggl\{ -8 \eta \del \Delta_0 \left(
 a \cdot b \tau^2 + a \cdot v \tau^3 +{1\over 3}{w \cdot b}\tau^3
\right)\Delta_0\nn\\
&&\qquad
- \sum_{\pm} \eta \left( \del \mp v\tau \right) \Delta_{\pm 2}
\left(a\cdot b\tau^2 + a\cdot v\tau^3 +{1\over 3}{w \cdot b}\tau^3
\pm 2 a \cdot v {\tau \over v} \right) \Delta_{\pm 2} \nn\\
&&\qquad
 + \sum_{\pm}  \mp \eta {1 \over v} \left(
a \cdot b \tau + \half a \cdot v \tau^2 + \half w \cdot b \tau^2 \right)
\Delta_{\pm 2}\biggr\} ,\\[5pt]
&&G_1 =-4\Tr\biggl\{ \eta \left( \Delta_0^2 \left(
a \cdot b + w \cdot b \tau \right) \del \Delta_0
-\left( a \cdot b + w \cdot b \tau \right) \del \Delta_0^3 \right) \nn\\
&&\qquad
+ \dot\eta\left( -\left( a \cdot b + w \cdot b \tau \right)
\Delta_0^3 + \del \Delta_0 \left( a \cdot b \tau +
\half w \cdot b \tau^2 \right) \Delta_0^2 - \left( a \cdot b \tau +
\half w \cdot b \tau^2 \right) \del \Delta_0^3 \right) \nn\\
&&\qquad
+ \ddot\eta\left( -\left( \half a \cdot b \tau
+ {1 \over 4} w \cdot b \tau^2 \right) \Delta_0^3 + \Delta_0 \left(
\half a \cdot b \tau + {1 \over 4} w \cdot b \tau^2 \right) \Delta_0^2
\right)\biggr\} . \label{G1_beyond_Tr}
\end{eqnarray}
Let us give a sample calculation of the first term in
(\ref{G1_beyond_Tr}) to explain how to evaluate it using the
convolution formulas:
\begin{eqnarray}
&&\int\!\d \tau\,\eta(\tau)
\bra{\tau}\Delta_0^2\del\Delta_0\ket{\tau}
= \int\!\d \tau\d\tau_0\, \eta(\tau)\int\!\d s_1\d s_2\d s_3
\calD(\tau,\tau_0;s_1+s_2)\,
\del_{\tau_0} \calD (\tau_0,\tau;s_3)e^{-b^2\sum s_k} \nn\\
&&
=\int\!\d\tau\,\eta(\tau)\!\int\!\d s_1 \d s_2 \d s_3\, v\,
\frac{-\cosh 2s_3 v\,\calH_1(\tau,\tau;s_1+s_2,s_3)+\tau}
{\sinh 2 s_3 v} \calD(\tau,\tau;\sum s_k)e^{-b^2\sum s_k} ,
\label{sample}
\end{eqnarray}
where $\calH_m$ is defined by
\begin{equation}
\calH_m(\tau_1,\tau_2;s_1,s_2)\calD(\tau_1,\tau_2;s_1+s_2)
\equiv\left.\left(\frac{\del}{\del J}\right)^m
\calG(\tau_1,\tau_2;s_1,s_2;J)\right|_{J=0} ,
\end{equation}
with $\calG$ of (\ref{convolution}).
In (\ref{sample}), we first applied (\ref{Delta_0^m}) to $\Delta_0^2$,
and then replaced $\tau_0$ arising from the $\tau_0$-derivative acting
on $\calD(\tau_0,\tau;s_3)$ with $\calH_1$.
Finally, we have to Taylor-expand (\ref{sample}) with respect to $v$
using the expression (\ref{keyformula}) and carry out the $s_k$
integrations.

After tedious but straightforward calculations, we get, keeping only
the terms with dimension $[\L^{-3}\T^{-3}]$,
\begin{eqnarray}
&&\Delta \Gamma_1 = \Delta \Gamma_1^{\rm eikonal}
+ \int\!\d\tau\,\eta\Biggl\{a\cdot b\left(
\frac{35}{4} \frac{v^2 \tau}{\rho^7}
-\frac{525}{16} \frac{v^4 \tau^3}{\rho^9}
+\frac{945}{32} \frac{v^6 \tau^5}{\rho^{11}}\right) \nn\\
&&\hspace*{3.5cm}
+ a \cdot v \left(
-\frac{3}{8} \frac{1}{\rho^5}
+\frac{135}{8} \frac{v^2 \tau^2}{\rho^7}
-\frac{1365}{32} \frac{v^4 \tau^4}{\rho^9}
+\frac{945}{32} \frac{v^6 \tau^6}{\rho^{11}} \right) \nn\\
&&\hspace*{3.5cm}
 + w \cdot b \left( -\frac{5}{8} \frac{1}{\rho^5}
+\frac{55}{8} \frac{v^2 \tau^2}{\rho^7}
-\frac{455}{32} \frac{v^4 \tau^4}{\rho^9}
+\frac{315}{32} \frac{v^6 \tau^6}{\rho^{11}}\right)\Biggr\} ,
\label{DeltaGamma1_beyond}\\[5pt]
&&G_1 = \int\!\d\tau
\Biggl\{a\cdot b \left(-\eta\frac{5}{2} \frac{v^2 \tau}{\rho^7}
+ \dot\eta\frac{1}{4}\frac{1}{\rho^5} \right)
+ w\cdot b\left(\eta\left[\frac{1}{2}\frac{1}{\rho^5}
-\frac{5}{2}\frac{v^2 \tau^2}{\rho^7} \right]
+ \dot\eta\frac{1}{4}\frac{\tau}{\rho^5} \right)\Biggr\} .
\label{G1_beyond}
\end{eqnarray}
Then, making the integration by parts for the $\dot\eta$ terms and
using
\begin{eqnarray}
\frac{\dddot{B} \cdot B}{B^5}\bb
&=&\bb\frac{w \cdot b}{\rho^5} ,\nn\\
\frac{\ddot{B} \cdot \dot{B}}{B^5}\bb
&=&\bb \frac{a \cdot v}{\rho^5} ,\nn\\
\frac{(\ddot{B} \cdot B)(\dot{B} \cdot B)}{B^7}
\bb &=&\bb \frac{a \cdot b\, v^2 \tau}{\rho^7}
+ \frac{a \cdot v\, v^2 \tau^2}{\rho^7}
+ \frac{w \cdot b\, v^2 \tau^2}{\rho^7} ,\nn\\
\frac{\dot{B}^2 (\dot{B} \cdot B)}{B^7}
\bb &=&\bb\frac{v^4 \tau}{\rho^7}
+ a\cdot b\left(\frac{v^2 \tau}{\rho^7}
- \frac{7}{2}\frac{v^4 \tau^3}{\rho^9}\right)
+ a\cdot v\left(\frac{7}{2}\frac{v^2 \tau^2}{\rho^7}
- \frac{7}{2}\frac{v^4 \tau^4}{\rho^9}\right) \nn\\
&&+ w \cdot b\left(\frac{1}{2}\frac{v^2 \tau^2}{\rho^7}
- \frac{7}{6}\frac{v^4\tau^4}{\rho^9}\right) ,\nn\\
\frac{(\dot{B} \cdot B)^3}{B^9}
\bb &=&\bb\frac{v^6 \tau^3}{\rho^9}
+ a\cdot b\left(3\frac{v^4\tau^3}{\rho^9}
-\frac{9}{2}\frac{v^6 \tau^5}{\rho^{11}}\right)
+ a\cdot v\left(\frac{9}{2}\frac{v^4\tau^4}{\rho^9}
-\frac{9}{2}\frac{v^6 \tau^6}{\rho^{11}}\right) \nn\\
&&+ w\cdot b\left(\frac{3}{2}\frac{v^4\tau^4}{\rho^9}
- \frac{3}{2}\frac{v^6\tau^6}{\rho^{11}}\right) ,
\end{eqnarray}
we can make the following identification for the sum of
(\ref{DeltaGamma1_beyond}) and (\ref{G1_beyond}):
\begin{eqnarray}
\int\!\d \tau\,\eta
\left( -\frac{3}{8} \frac{\dddot{B} \cdot B}{B^5}
-\frac{3}{8} \frac{\ddot{B} \cdot \dot{B}}{B^5}
+\frac{15}{4} \frac{(\ddot{B} \cdot B)(\dot{B} \cdot B)}{B^7}
+\frac{15}{4} \frac{\dot{B}^2 (\dot{B} \cdot B)}{B^7}
-\frac{105}{16} \frac{(\dot{B} \cdot B)^3}{B^9} \right) .
\end{eqnarray}
Using the same method we can calculate the $\eta$-independent terms
in $\Goneloop$. This time there are 11 possibilities such as
$\ddddot{B}\cdot B/B^5$, $\dddot{B}\cdot\dot{B}/B^5$,
$\ddot{B}\cdot\ddot{B}/B^5$, $(\dddot{B}\cdot B)(\dot{B}\cdot B)/B^7$,
$\dots$, $(\dot{B}\cdot B)^4/B^{11}$.
To determine their coefficients we have to expand $B_m$ as
$B_m = b_m + v_m \tau + a_m \tau^2 /2 + w_m \tau^3/6 + z_m \tau^4/24$
and keep terms linear in $a\cdot b$, $a\cdot v$, $a^2$, $w\cdot b$,
$w\cdot v$ and $z\cdot b$.
Surprisingly, however, the $\eta$-independent terms of $\Goneloop$
remains the same as (\ref{final_Gamma1_eikonal}) obtained in the
eikonal approximation.

Summing up the results of our calculations, we conclude that the total
1-loop effective action is given by
\begin{equation}
\Goneloop
= \int\!\d\tau\left\{ -\frac{15}{16}\frac{(\dot{B}^2)^2}{B^7}
+\eta\left(
\frac{15}{8} \frac{\dot{B}^2 (\dot{B} \cdot B)}{B^7}
+ \frac{\d}{\d\tau}\! \left[-\frac{3}{2}\frac{1}{B}
+\frac{15}{16}\frac{(\dot{B} \cdot B)^2}{B^7}
- \frac{3}{8}\frac{\ddot{B}\cdot B}{B^5}\right]\right)\right\} .
\end{equation}
This has the desired coefficient $15/8$ and agrees essentially with
the claim (\ref{JKYclaim}) of \cite{JKY0}, implying that the SCT
Ward identity is in fact satisfied.

\section{Conclusion}

In this paper we calculated the 1-loop effective action in Matrix
model beyond the eikonal approximation and confirmed the SCT Ward
identity. Eikonal approximation can only give limited information on
the effective action. It is crucial to consider a higher
polynomial form for $B_m(\tau)$ for determining completely
the effective action, especially, its $\eta$-dependent terms.
We finish this paper by giving some comments.

\noindent
$\bullet$ In our calculation, it is important to take into account the
contribution of $G_1$ (\ref{G1}), which vanishes in the eikonal
approximation and hence is usually dropped from the start.

\noindent
$\bullet$ By applying our method of calculation, it is possible to
determine also the higher order terms with dimensions
$[\L^{-5}\T^{-6}]$, $[\L^{-7}\T^{-8}]$, $\ldots$, in the derivative
expansion of the effective action (cf.\ eq.\ (\ref{expansion})).

\noindent
$\bullet$ In this paper we examined the SCT Ward identity (\ref{WI})
at the lowest order in $\eta$. By keeping terms with higher powers of
$\eta$,  the SCT Ward identity is expected to hold without putting
$\eta=0$ after the transformation.

\vspace{.7cm}
\noindent
{\large\bf Acknowledgments}\\[.2cm]
We would like to thank T.\ Yoneya for his enlightening lecture at
Department of Physics, Kyoto University.
One of the authors (S.\ M.) would like to thank Y.\  Kazama for his
clear and self-contained lecture at Ochanomizu University.

\newcommand{\J}[4]{{\sl #1} {\bf #2} (#3) #4}
\newcommand{\andJ}[3]{{\bf #1} (#2) #3}
\newcommand{\AP}{Ann.\ Phys.\ (N.Y.)}
\newcommand{\MPL}{Mod.\ Phys.\ Lett.}
\newcommand{\NP}{Nucl.\ Phys.}
\newcommand{\PL}{Phys.\ Lett.}
\newcommand{\PR}{Phys.\ Rev.}
\newcommand{\PRL}{Phys.\ Rev.\ Lett.}
\newcommand{\ATMP}{Adv.\ Theor.\ Math.\ Phys.}

\end{document}